\documentclass[zpreprint,zbstdefault]{./zeus_paper}

\usepackage[english]{babel}
\usepackage{url}

\newcommand{\ZcoosysB}{%
The ZEUS coordinate system is a right-handed Cartesian system, with the $Z$
axis pointing in the proton beam direction, referred to as the ``forward
direction'', and the $X$ axis pointing left towards the centre of HERA.
The coordinate origin is at the nominal interaction point.\xspace}
\newcommand{\ZcoosysfnBeta}{\footnote{\ZcoosysB}} 
\newcommand{\Zdetdesc}{%
A detailed description of the ZEUS detector can be found 
elsewhere~\cite{zeus:1993:bluebook}. A brief outline of the 
components that are most relevant for this analysis is given
below.\xspace}
\newcommand{\Zctddesc}[1]{%
Charged particles are tracked in the central tracking detector (CTD)~\citeCTD,
which operates in a magnetic field of $1.43\Tesla$ provided by a thin 
superconducting solenoid. The CTD consists of 72~cylindrical drift chamber 
layers, organised in nine superlayers covering the polar-angle#1 region 
\mbox{$15^\circ<\theta<164^\circ$}. The transverse-momentum resolution for
full-length tracks is $\sigma(p_T)/p_T=0.0058p_T\oplus0.0065\oplus0.0014/p_T$,
with $p_T$ in $\Gev$.}
\newcommand{\Zcaldesc}{%
The high-resolution uranium--scintillator calorimeter (CAL)~\citeCAL consists 
of three parts: the forward (FCAL), the barrel (BCAL) and the rear (RCAL)
calorimeters. Each part is subdivided transversely into towers and
longitudinally into one electromagnetic section (EMC) and either one (in RCAL)
or two (in BCAL and FCAL) hadronic sections (HAC). The smallest subdivision of
the calorimeter is called a cell.  The CAL energy resolutions, as measured under
test-beam conditions, are $\sigma(E)/E=0.18/\sqrt{E}$ for electrons and
$\sigma(E)/E=0.35/\sqrt{E}$ for hadrons, with $E$ in $\Gev$.}

\chardef\usc=95
\chardef\til=126
\catcode`\@=11 
\DeclareRobustCommand\xdotspace{\futurelet\@let@token\@xdotspace}
\def\@xdotspace{%
  \ifx\@let@token.\else
  \ifx\@let@token\bgroup.\else
  \ifx\@let@token\egroup.\else
  \ifx\@let@token\/.\else
  \ifx\@let@token\ .\else
  \ifx\@let@token~.\else
  \ifx\@let@token!.\else
  \ifx\@let@token,.\else
  \ifx\@let@token:.\else
  \ifx\@let@token;.\else
  \ifx\@let@token?.\else
  \ifx\@let@token/.\else
  \ifx\@let@token'.\else
  \ifx\@let@token).\else
  \ifx\@let@token-.\else
  \ifx\@let@token\@xobeysp.\else
  \ifx\@let@token\space.\else
  \ifx\@let@token\@sptoken.\else
   .\space
   \fi\fi\fi\fi\fi\fi\fi\fi\fi\fi\fi\fi\fi\fi\fi\fi\fi\fi}
\catcode`\@=12 

\newcommand{\stru}[2]{%
   \relax\ifmmode\hbox{\vrule height#1 depth#2 width0pt}%
   \else\vrule height#1 depth#2 width0pt\fi}

\newcommand{\Ronum}[1]{\uppercase\expandafter{\romannumeral#1}}
\newcommand{\ronum}[1]{\expandafter{\romannumeral#1}}
\DeclareRobustCommand{\LaTeXZ}{%
  \LaTeX\kern-.05em4\kern-.1em
  {\raisebox{-0.2ex}{$\scriptstyle\text{ZEUS}$}}\xspace}

\DeclareMathAlphabet{\mathbf}{OT1}{cmr}{bx}{sl}
\newcommand{\eVdist}{\kern-0.06667em}

\newcommand{\Gev}{{\text{Ge}\eVdist\text{V\/}}}

\newcommand{\gev}{{\,\text{Ge}\eVdist\text{V\/}}}

\newcommand{\cm}{\,\text{cm}}

\newcommand{\Tesla}{\,\text{T}}

\newcommand{\slashfrac}[2]{%
  \raisebox{0.5ex}{\ensuremath #1}\kern-0.12em/\kern-0.08em
  \raisebox{-.8ex}{\ensuremath #2}}

\newcommand{\sqr}[3]{%
    {\vcenter{\hrule height.#3ex\hbox{\vrule width.#2ex height#1ex
     \kern#1ex\vrule width.#3ex}\hrule height.#2ex}}}

\catcode`\@=11 
\newcommand{\parenbar}{\mathpalette\p@renb@r}
\def\p@renb@r#1#2{\vbox{%
  \ifx#1\scriptscriptstyle \dimen@.7em\dimen@ii.2em\else
  \ifx#1\scriptstyle \dimen@.8em\dimen@ii.25em\else
  \dimen@1em\dimen@ii.4em\fi\fi \offinterlineskip
  \ialign{\hfill##\hfill\cr
    \vbox{\hrule width\dimen@ii}\cr
    \noalign{\vskip-.3ex}%
    \hbox to\dimen@{$\mathchar300\hfil\mathchar301$}\cr
    \noalign{\vskip-.3ex}%
    $#1#2$\cr}}}
\catcode`\@=12 

\newcommand{\IP}{{\rm I$\kern-0.01667em$P}\xspace}

\mathchardef\qsm=63
\mathchardef\pls=43
\mathchardef\mns=512
\mathchardef\plm=518
\mathchardef\eql=61
\mathchardef\smallleft=300
\mathchardef\smallright=301
\mathchardef\les=316
\mathchardef\gre=318
\mathchardef\leq=532
\mathchardef\grq=533

\catcode`\@=11 
\newcounter{pict@width}
\newcounter{pict@height}
\newlength{\pict@scale}
\newcommand{\psfigadd}[4]{%
\setcounter{pict@width}{1*\ratio{#2+\pict@scale/2}{\pict@scale}}
\setcounter{pict@height}{1*\ratio{#3+\pict@scale/2}{\pict@scale}}
\setlength{\unitlength}{\pict@scale}
\hbox to #2{\hspace{-\fill}\begin{picture}(\thepict@width,\thepict@height)
\put(0,0){\psfig{figure=#1,width=#2,height=#3,clip=}}
\SetScale{0.283466457}
\SetWidth{1.763889}
{#4}
\end{picture}}
}
\newcounter{pict@widthfst}
\newcounter{pict@widthscd}
\newcounter{pict@widthtot}
\newcommand{\psfigaddtwo}[7]{%
\setcounter{pict@widthfst}{1*\ratio{#2+\pict@scale/2}{\pict@scale}}
\setcounter{pict@widthscd}{1*\ratio{#2+#4+\pict@scale/2}{\pict@scale}}
\setcounter{pict@widthtot}{1*\ratio{#2+#4+#6+\pict@scale/2}{\pict@scale}}
\setcounter{pict@height}{1*\ratio{#3+\pict@scale/2}{\pict@scale}}
\setlength{\unitlength}{\pict@scale}
\hbox{\hspace{-\fill}\begin{picture}(\thepict@widthtot,\thepict@height)
\put(0,0){\psfig{figure=#1,width=#2,height=#3,clip=}}
\put(\thepict@widthscd,0){\psfig{figure=#5,width=#6,height=#3,clip=}}
\SetScale{0.283466457}
\SetWidth{1.763889}
{#7}
\end{picture}}
}
\newcommand{\psfigror}[4]{%
\setcounter{pict@width}{1*\ratio{#2+\pict@scale/2}{\pict@scale}}
\setcounter{pict@height}{1*\ratio{#3+\pict@scale/2}{\pict@scale}}
\setlength{\unitlength}{\pict@scale}
\hbox{\begin{picture}(\thepict@width,\thepict@height)
\put(0,\thepict@height){\psfig{figure=#1,width=#3,height=#2,clip=,angle=270}}
\SetScale{0.283466457}
\SetWidth{1.763889}
{#4}
\end{picture}}
}
\newcommand{\psfigrol}[4]{%
\setcounter{pict@width}{1*\ratio{#2+\pict@scale/2}{\pict@scale}}
\setcounter{pict@height}{1*\ratio{#3+\pict@scale/2}{\pict@scale}}
\setlength{\unitlength}{\pict@scale}
\hbox{\begin{picture}(\thepict@width,\thepict@height)
\put(0,0){\psfig{figure=#1,width=#3,height=#2,clip=,angle=90}}
\SetScale{0.283466457}
\SetWidth{1.763889}
{#4}
\end{picture}}
}
\catcode`\@=12 
\newlength\listtextwidth

\catcode`\@=11 
\newlength{\@tabfninsert}
\newlength{\@tabfnwidth}
\newcommand{\tabfootnote}[2]{%
  \setlength{\@tabfninsert}{0.8em}
  \setlength{\@tabfnwidth}{\textwidth}
  \addtolength{\@tabfnwidth}{-\@tabfninsert}
  \addtolength{\@tabfnwidth}{-0.4em}
  \noindent\makebox[\@tabfninsert][r]{\footnotesize$^{#1}$\hfil}\hfill%
  \parbox[t]{\@tabfnwidth}{\footnotesize #2\hfill}}
\catcode`\@=12 

\newcommand{\ud}{\mathrm{d}}
\def\citeCTD{{\cite{%
nim:a279:290,*npps:b32:181,*nim:a338:254%
}}\xspace}
\def\citeCAL{{\cite{%
nim:a309:77,*nim:a309:101,*nim:a321:356,*nim:a336:23%
}}\xspace}

\begin{document}
\prepnum{DESY 07-070}

\title{
Measurement of (anti)deuteron and (anti)proton production
in DIS at HERA 
}                                                       
                    
\author{ZEUS Collaboration}
\date{May 24, 2007}

\abstract{
The first observation of (anti)deuterons in deep 
inelastic scattering at HERA has been
made with the ZEUS detector at a centre-of-mass energy of 300--318 $\gev$
using an integrated luminosity of $120$~pb$^{-1}$.
The measurement was  performed in the central rapidity region
for transverse momentum 
per unit of mass in the range $0.3<p_T/M<0.7$. 
The particle rates have been extracted and interpreted in terms of the coalescence model.
The (anti)deuteron production yield 
is smaller than the (anti)proton yield by approximately three orders of magnitude, 
consistent with the world measurements.
}
\makezeustitle

\pagenumbering{Roman}                                                                              
\begin{center}                                                                                     
{                      \Large  The ZEUS Collaboration              }                               
\end{center}                                                                                       
  S.~Chekanov$^{   1}$,                                                                            
  M.~Derrick,                                                                                      
  S.~Magill,                                                                                       
  B.~Musgrave,                                                                                     
  D.~Nicholass$^{   2}$,                                                                           
  \mbox{J.~Repond},                                                                                
  R.~Yoshida\\                                                                                     
 {\it Argonne National Laboratory, Argonne, Illinois 60439-4815}, USA~$^{n}$                       
\par \filbreak                                                                                     
  M.C.K.~Mattingly \\                                                                              
 {\it Andrews University, Berrien Springs, Michigan 49104-0380}, USA                               
\par \filbreak                                                                                     
  M.~Jechow, N.~Pavel~$^{\dagger}$, A.G.~Yag\"ues Molina \\                                        
  {\it Institut f\"ur Physik der Humboldt-Universit\"at zu Berlin,                                 
           Berlin, Germany}
% ...
\newpage                                                                                           
$^{\    1}$ supported by DESY, Germany \\                                                          
$^{\    2}$ also affiliated with University College London, UK \\                                  
% ...
$^{\dagger}$ deceased \\                                                                           
%
\newpage   
                                                           %
                                                           %
\begin{tabular}[h]{rp{14cm}}                                                                       
$^{a}$ &  supported by the Natural Sciences and Engineering Research Council of Canada (NSERC) \\  
$^{b}$ &  supported by the German Federal Ministry for Education and Research (BMBF), under        
          contract numbers HZ1GUA 2, HZ1GUB 0, HZ1PDA 5, HZ1VFA 5\\                                
$^{c}$ &  supported in part by the MINERVA Gesellschaft f\"ur Forschung GmbH, the Israel Science   
          Foundation (grant no. 293/02-11.2) and the U.S.-Israel Binational Science Foundation \\  
$^{d}$ &  supported by the German-Israeli Foundation and the Israel Science Foundation\\           
$^{e}$ &  supported by the Italian National Institute for Nuclear Physics (INFN) \\                
$^{f}$ &  supported by the Japanese Ministry of Education, Culture, Sports, Science and Technology 
          (MEXT) and its grants for Scientific Research\\                                          
$^{g}$ &  supported by the Korean Ministry of Education and Korea Science and Engineering          
          Foundation\\                                                                             
$^{h}$ &  supported by the Netherlands Foundation for Research on Matter (FOM)\\                   
$^{i}$ &  supported by the Polish State Committee for Scientific Research, grant no.               
          620/E-77/SPB/DESY/P-03/DZ 117/2003-2005 and grant no. 1P03B07427/2004-2006\\             
$^{j}$ &  partially supported by the German Federal Ministry for Education and Research (BMBF)\\   
$^{k}$ &  supported by RF Presidential grant N 8122.2006.2 for the leading                         
          scientific schools and by the Russian Ministry of Education and Science through its grant
          Research on High Energy Physics\\                                                        
$^{l}$ &  supported by the Spanish Ministry of Education and Science through funds provided by     
          CICYT\\                                                                                  
$^{m}$ &  supported by the Particle Physics and Astronomy Research Council, UK\\                   
$^{n}$ &  supported by the US Department of Energy\\                                               
$^{o}$ &  supported by the US National Science Foundation. Any opinion,                            
findings and conclusions or recommendations expressed in this material                             
are those of the authors and do not necessarily reflect the views of the                           
National Science Foundation.\\                                                                     
$^{p}$ &  supported by the Polish Ministry of Science and Higher Education                         
as a scientific project (2006-2008)\\                                                              
$^{q}$ &  supported by FNRS and its associated funds (IISN and FRIA) and by an Inter-University    
          Attraction Poles Programme subsidised by the Belgian Federal Science Policy Office\\     
$^{r}$ &  supported by the Malaysian Ministry of Science, Technology and                           
Innovation/Akademi Sains Malaysia grant SAGA 66-02-03-0048\\                                       
\end{tabular}                                                                                      
                                                           %
                                                           %

\pagenumbering{arabic}
\pagestyle{plain}
\section{Introduction}
\label{sec-int}

Light stable nuclei, such as deuterons ($d$) and tritons ($t$),
are loosely bound states whose
production mechanism in high-energy collisions is poorly understood.
Most measurements of light stable nuclei have been performed for antideuterons ($\bar{d}$).
A selection of $d$ from primary interactions
is more difficult as it requires separation of such states from 
particles produced by
interactions of colliding beams with residual gas in the beam pipe and
by secondary interactions in detector material.
The first observation of $\bar{d}$  \cite{first_obs} was followed by
a number of experiments on antideuteron production.
The production rate of $\bar{d}$  in $e^{+}e^{-}\to q\bar{q} $ collisions
\cite{argus1,*argus2,opal1,aleph_deu,cleo_deu}
is significantly lower than that measured in
$\Upsilon (1S)$ and $\Upsilon (2S)$ decays \cite{argus1,*argus2,cleo_deu}.
The $\bar{d}$ rate in $e^{+}e^{-}\to  q\bar{q} $ is also lower
than that in proton-nucleus ($pA$) \cite{pA1,*pA2,pA3}, proton-proton
($pp$) \cite{pp1,*pp2,*Abramov:1986ti} and
photon-proton  ($\gamma p$) collisions at HERA~\cite{h1deuterons}, but
higher than that in nucleus-nucleus  collisions
\cite{aa1,*aa2,*star,*Ahle:1998jv,*Bearden:1999iq,*Bearden:2002ta,phenix1}.
For heavy-ion collisions, the coalescence model~\cite{cmodel} was proposed to explain the
production of $d$($\bar{d}$).

This paper presents the results of the first measurement
of $d$ and $\bar{d}$
in the central rapidity region of
deep inelastic $ep$ scattering (DIS). 
The analysis was performed for exchanged photon virtuality, $Q^2$, above 1 $\gev^2$. 

\section{Coalescence model for (anti)deuteron formation}
\label{sec-dec}

According to the coalescence model~\cite{cmodel} developed for heavy-ion collisions,
the production rate of $d$ is determined by the overlap between the 
wave-function of a proton ($p$) and a neutron ($n$)
with the wave-function of a $d$.
In this case, the  $d$  cross section is the product of 
single-particle cross sections for protons and neutrons, with a coefficient of proportionality
reflecting the spatial size of the fragmentation region emitting the particles.
The same approach applies for  $\bar{d}$ production.
This model was also used to describe $d(\bar{d})$ production
in  $pp$~\cite{pp1,*pp2,*Abramov:1986ti}, $\gamma p$~\cite{h1deuterons}
and $e^+e^-$ \cite{argus1,*argus2,aleph_deu} interactions.

Assuming that all baryons are uncorrelated and the invariant
differential cross section for neutrons is equal to that for protons,
the invariant differential cross section for deuteron
formation can be parameterised as
$$
\frac{E_d}{\sigma_{tot}} \frac{\ud^3 \sigma_d} {\ud p_d^3} = B_2
\left(  \frac{E_p}{\sigma_{tot}}    \frac{\ud^3 \sigma_p}{\ud p_p^3} \right)^2,
$$
where  $E_{d(p)}$ and $\sigma_{d(p)}$ are the energy and the production
cross section of  the $d$($p$), respectively,
$p_d(p_p)$ is the momentum of the $d$($p$)  and
$\sigma_{tot}$ is the total $ep$ cross section for the considered kinematic range.
The coalescence parameter, $B_2$, is inversely proportional 
to the volume of the fragmentation region emitting the particles.
The same relation holds for $\bar{d}$  and  $\bar{p}$.
If $B_2$ is the same for particles and antiparticles,
then the production ratio $\bar{d}/d$ is equal to $(\bar{p}/p)^2$.
The coalescence parameter can be obtained from 
$$
B_2 =  \Biggl(\frac{E_d}{\sigma_{tot}} \frac{\ud^3\sigma_d}{\ud p_d^3}\Biggl)
       \Biggl(\frac{E_p}{\sigma_{tot}} \frac{\ud^3\sigma_p}{\ud p_p^3}\Biggl)^{-2}
   = M^4_p \,   M^{-2}_d\,   R^2 (d/p)\,
\left(  \frac{\gamma_d}{\sigma_{tot}}    \frac{\ud^3 \sigma_d}{\ud (p_d/M_d)^3} \right)^{-1},
$$
where $M_{d(p)}$ is the mass of the $d$($p$), $\gamma_d=E_d/M_d$,
$R(d/p)$ is the ratio of the number of $d$ to $p$ expressed as a function of $p_T/M_{d(p)}$,  with $p_T$ being
the transverse momentum~\cite{h1deuterons}.

\section{Experimental set-up}
\label{sec-exp}

\Zdetdesc

\Zctddesc\ZcoosysfnBeta\
To estimate the ionisation energy loss per unit length, $dE/dx$, of
particles in the CTD\cite{pl:b481:213,*epj:c18:625,*thesis:dedx}, the truncated
mean of the anode-wire pulse heights was calculated, which removes the
lowest $10\%$ and at least the highest $30\%$ depending on the number
of saturated hits. The measured $dE/dx$ values were corrected by
normalising to the average $dE/dx$ for tracks around the region of
minimum ionisation for pions with momentum $p$ satisfying 
$0.3~<~p~<~0.4\gev$. Henceforth, $dE/dx$
is quoted in units of minimum ionising particles (mips).
The resolution of the  $dE/dx$ measurement for full-length
tracks is about $9\%$.

\Zcaldesc
$\>$ A presampler \cite{nim:a382:419,*magill:bpre} mounted in front of
the calorimeter 
and a scintillator-strip detector (SRTD)~\cite{nim:a401:63}
were used to correct the energy of the scattered
electron\footnote{Henceforth the term
electron is used to refer both to electrons and positrons.}.
The position of electrons scattered
close to the electron beam direction is determined by the SRTD
detector.

The inactive material between the interaction region and the CTD, 
relevant for this analysis,  consists of the central
beam pipe made of aluminum with  $1.5$~mm wall thickness 
and the inner diameter of $135$~mm.
The CTD inner wall with a diameter of $324$~mm consists of 
two aluminum skins, each  $0.7$~mm thick, separated  
by a $8.6$~mm gap filled with polyurethane foam with a nominal density of 0.05~g/cm$^3$.

The luminosity was measured using the bremsstrahlung process 
$ep \to ep \gamma$ with the luminosity
monitor~\cite{Desy-92-066,*zfp:c63:391,*acpp:b32:2025}, a
lead--scintillator calorimeter placed in the HERA tunnel at $Z = -107$~m.

\section{Monte Carlo simulation}
\label{sec:mc}

To study the detector response, the {\sc Ariadne} 4.12  Monte Carlo (MC)
model~\cite{cpc:71:15} for the description of inclusive DIS events was used.
The {\sc Ariadne}  program uses the Lund string model
\cite{prep:97:31} for hadronisation, as implemented in
{\sc Pythia} 6.2\cite{cpc:46:43,cpc:82:74,cpc:135:238}.
In its original version, this MC  does not include a mechanism for the production of $d$
or  other light stable nuclei.
To determine reconstruction efficiencies, a second {\sc Ariadne} sample was
generated in which $d$'s were included at the generator level by combining 
$p$ and $n$ with similar momenta.

The {\sc Ariadne} events
were passed through a full simulation of the
detector using the {\sc Geant} 3.13 \cite{tech:cern-dd-ee-84-1} program.
The {\sc Geant} simulation uses the {\sc Gheisha} model \cite{Fesefeldt:1985yw} to simulate
hadronic interactions in the material.
The {\sc Geant}  program  cannot be used for $\bar{d}$  as this particle
is not included in the particle table.

\section{Event sample}
\label{sec:data}

\subsection{DIS event selection}

The data sample
corresponds to an integrated luminosity of 120.3 pb$^{-1}$
taken between 1996 and 2000 with the ZEUS detector at HERA.
This sample consists of 38.6 pb$^{-1}$ of
$e^+p$ data taken at a centre-of-mass energy of $300 \gev$,  $65.0$
pb$^{-1}$ taken at $318
\gev$ and $16.7$  pb$^{-1}$ of $e^-p$ data taken at $318 \gev$.

The search was performed using DIS events with
exchanged-photon virtuality
$Q^2 > 1\gev^2$.
The event selection was similar to that used in a  previous ZEUS
publication~\cite{Chekanov:2004kn}.
A three-level trigger
\cite{zeus:1993:bluebook} was used to select events online.
At the third-level trigger, an 
electron with an energy greater than $4\gev$
was required.
Data below $Q^2 \approx 20~\gev^2$
were  prescaled to reduce trigger rates.

The Bjorken scaling variable, $x_{\mathrm{Bj}}$, and  $Q^2$ were
reconstructed using the electron method (denoted by the subscript
$e$), which uses measurements of the energy and angle of the scattered
electron.
The scattered-electron
candidate was identified from the pattern of energy deposits in the
CAL \cite{nim:a365:508}.
In addition, the inelasticity was reconstructed
using the Jacquet-Blondel
method\cite{proc:epfacility:1979:391}, $y_{\mathrm{JB}}$,  or the electron method, $y_e$.

For the final DIS sample, the  following requirements were imposed:

\begin{itemize}

\item[$\bullet$]
$Q_e^2 > 1\gev^2$;

\item[$\bullet$]
the impact point of the scattered electron on the RCAL outside the ($X, Y$) region
($\pm 12$, $\pm 6$) cm centred  on the beamline;

\item[$\bullet$]
$E_{e^{'}} >  8.5\gev$, where $E_{e^{'}}$ is the
energy of the scattered electron measured in the CAL and corrected for energy losses;

\item[$\bullet$]
$35\> < \>   \delta \> < \> 65$ GeV, where
$\delta=\sum E_i(1-\cos\theta_i)$,
$E_i$ is the energy of the $i$-th calorimeter
cell, $\theta_i$ is its polar angle
and the sum runs over all cells;

\item[$\bullet$]
$y_{e}\> < \> 0.95$ and $y_{\mathrm{JB}}\> > \> 0.01$;

\item[$\bullet$]
at least three tracks fitted to the primary vertex
to ensure a good reconstruction of the primary vertex and to
reduce contributions from non-$ep$ events;

\item[$\bullet$]
$\mid Z_{\mathrm{vtx}} \mid  <   40\cm$ and 
$\sqrt{ X_{\mathrm{vtx}}^2 + Y_{\mathrm{vtx}}^2} < 1\cm$,  where
$Z_{\mathrm{vtx}}$, $X_{\mathrm{vtx}}$ and $Y_{\mathrm{vtx}}$ are the coordinates
of the  vertex position
determined from the tracks.

\end{itemize}

The average $Q^2$ of  the selected sample was about 10 GeV$^2$.

\subsection{Track selection and the $\mathbf{dE/dx}$ measurement}

The present analysis is based on charged tracks measured in the
CTD. The tracks were required to have: 

\begin{itemize}

\item[$\bullet$]
at least 40  CTD
hits, with at least $8$ of them for the $dE/dx$ measurement;

\item[$\bullet$]
the transverse momentum $p_T\ge 0.15\gev$.

\end{itemize}

These cuts selected 
a region where the CTD track acceptance, as well as 
the resolutions in momentum and the $dE/dx$, were high.

To identify particles originating from  $ep$ collisions,
the following additional variables were  reconstructed for each track:

\begin{itemize}
\item[$\bullet$]
the distance,  $\Delta Z$,  of the $Z$-component of the
track helix to $Z_{\mathrm{vtx}}$;

\item[$\bullet$]
the distance of closest approach ($DCA$) of the track to the beam-spot location
in the transverse plane.
The beam-spot position is determined
from the average primary-vertex distributions in $X$ and $Y$  for each 
data-taking period.
The $DCA$ 
is assigned a positive (negative) value if the beam spot
lies left (right) of the particle path.
\end{itemize}

\begin{figure}
\begin{center}

\includegraphics[height=9.8cm]{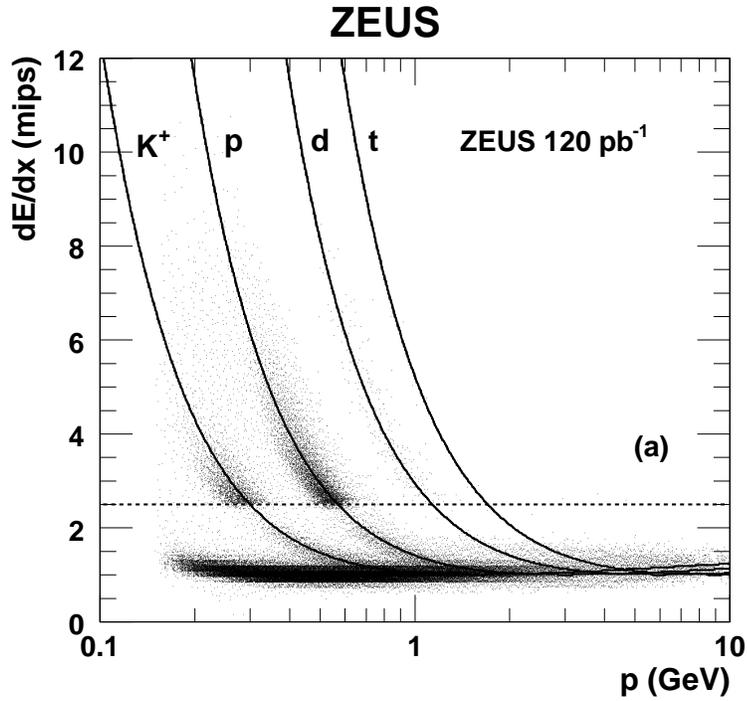}

\includegraphics[height=9.8cm]{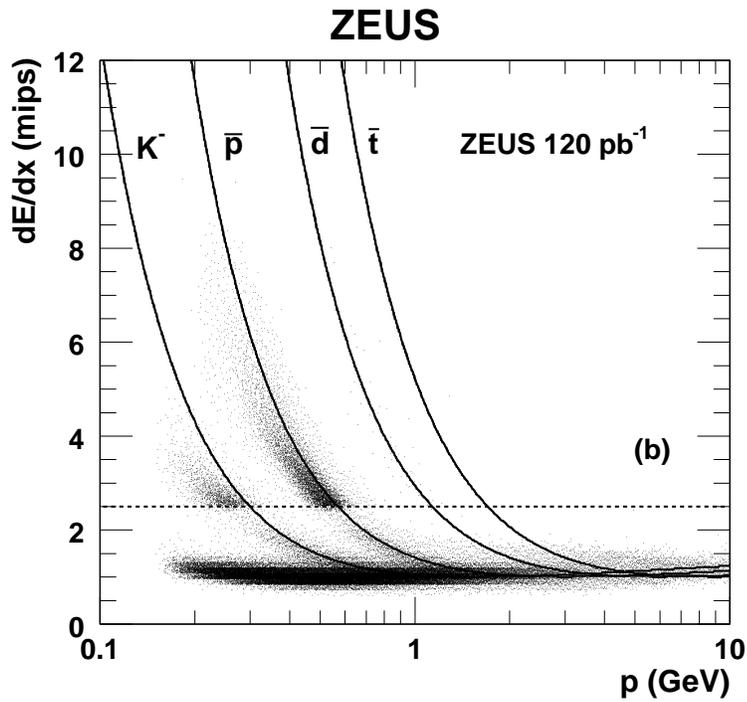}

\caption{The $dE/dx$ distributions  as a function of the track momentum
for (a) positive and (b) negative tracks. 
The DIS events were accepted by requiring 
at least one track with $dE/dx > 2.5$ mips (denoted by the dashed lines),  
$|\Delta Z|<1\cm$  and $|DCA|<0.5\cm$. 
The lines show the most-probable energy loss calculated using the
Bethe-Bloch formula for different particle
species.
}
\label{dedx}
\end{center}
\end{figure}

Figure~\ref{dedx} shows the $dE/dx$
distribution as a function of the track momentum for positive and
negative tracks.
The events were selected  by requiring at least one track
with  $dE/dx > 2.5$ mips.
To reduce the fraction of tracks coming from non-$ep$ collisions,
the tracks were  required to have $| \Delta Z | < 1\cm$ and $| DCA | <0.5\cm$.
After such a selection, clear
bands corresponding to charged kaons, protons and deuterons were observed.
The requirement $dE/dx > 2.5$~mips enhances the fraction of events with at least one
particle with a mass larger than the pion mass and leads to the discontinuity near
$dE/dx=2.5$ mips seen in Figure~\ref{dedx}.
The lines show the most probable energy loss calculated from the 
Bethe-Bloch formula~\cite{Yao:2006px}.
The $dE/dx$ bands  for $K^-$ and $\bar{p}$
are slightly shifted with respect to the Bethe-Bloch expectations due to the  
geometrical structure of the CTD drift cells which  leads to  a different
response to negative and positive tracks.

\begin{figure}
\begin{center}
  \includegraphics[height=16.0cm]{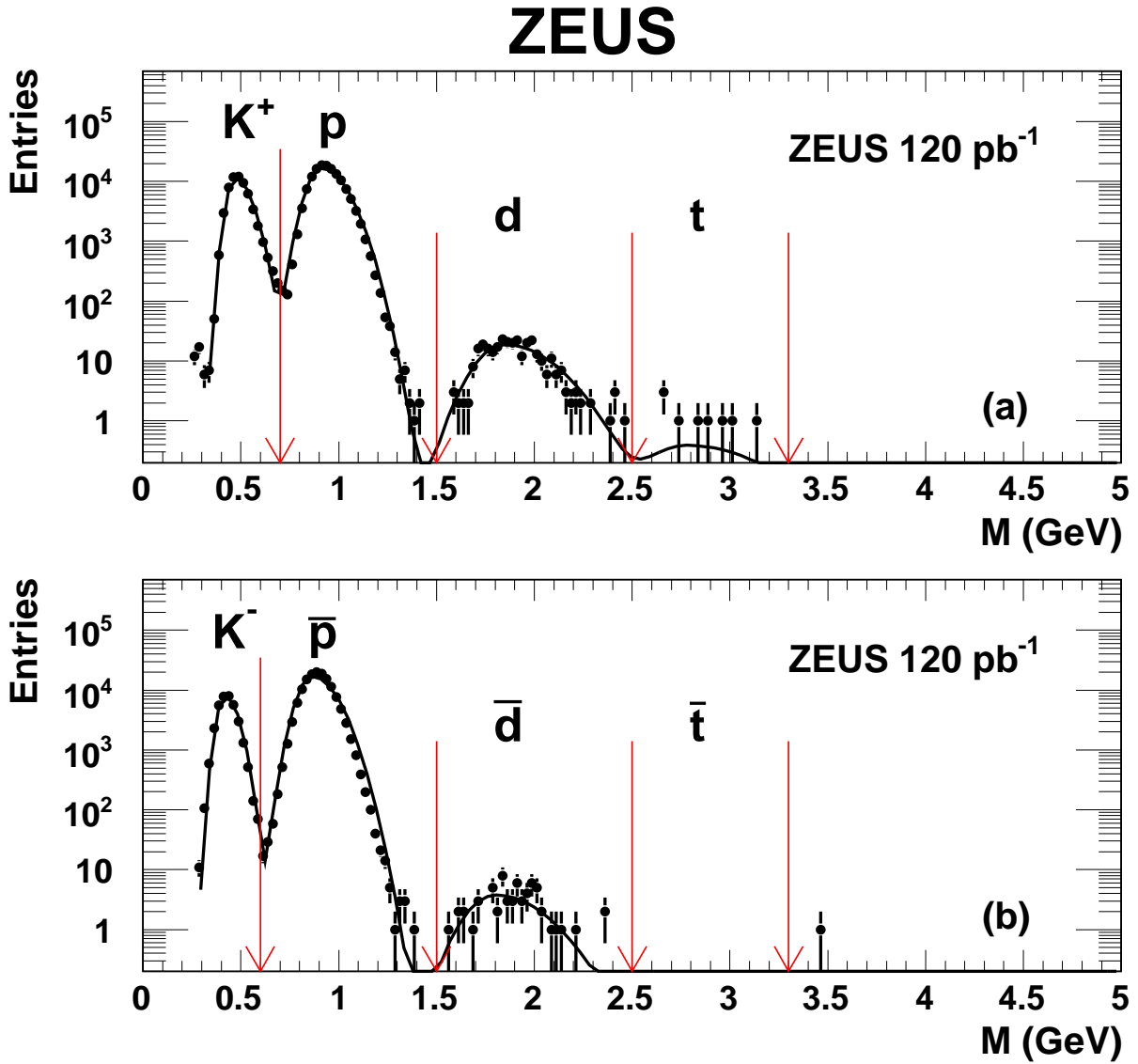}
\caption{The mass spectra for (a) positive and (b) negative particles.
Tracks are selected as for Figure~\ref{dedx}. 
The mass distribution was calculated from the track momenta and the $dE/dx$.  
The arrows indicate the cuts applied for the selection of candidates.
}
\label{mass}
\end{center}
\end{figure}

Figure~\ref{mass} shows the reconstructed masses, $M$,  for different particle species.
The masses  were  calculated from the measured track
momentum and energy loss using the Bethe-Bloch formula.
The mass distributions were fitted with
asymmetric\footnote{An asymmetric Gaussian has different
widths for the left and right parts of the function.}  Gaussian
functions.  The relative width  obtained was  $11\%$  
($7\%$)  for the left (right) part of the function.

The number of $p$($\bar{p}$) candidates in the mass region
$0.7(0.6)<M<1.5\gev$ was $1.61\times 10^5$ ($1.66\times 10^5$).
Due to a shift in the $dE/dx$ for negative tracks,
the lower mass cut for $\bar{p}$ was at $0.6\gev$.
The numbers  of $d$  and $\bar{d}$
in the mass window $1.5<M<2.5\gev$ were $309$ and $62$, respectively.
The number  of $p$  migrating to the $d$
mass region was estimated to be less than $1\%$ of
the total number of $d$
candidates. A similar estimate was obtained for antiparticles.
A small number of triton candidates was observed in the mass
window $2.5<M<3.5\gev$.
However, due  to low statistics,
it was difficult to establish a peak inside this mass window,
therefore, no conclusive statement
on the  origin of the tracks in the region $2.5<M<3.5\gev$ was possible.

The observed $p$($\bar{p}$)  and   $d$($\bar{d}$)
candidates were required to
be in the central rapidity region, $|y| < 0.4$, and to have
$0.3 < p_T/M < 0.7$.
This determines the kinematic range used for the cross-section calculations.

\subsection{Identification of particles produced in $ep$ collisions}
\label{sec:identif}

The observed $p$($\bar{p}$)  and   $d$($\bar{d}$)  candidates selected after the $dE/dx$ mass cuts
can originate from  secondary interactions
in the inactive material between the interaction point and
the central tracking detector.

\begin{figure}
\begin{center}
  \includegraphics[height=16.0cm]{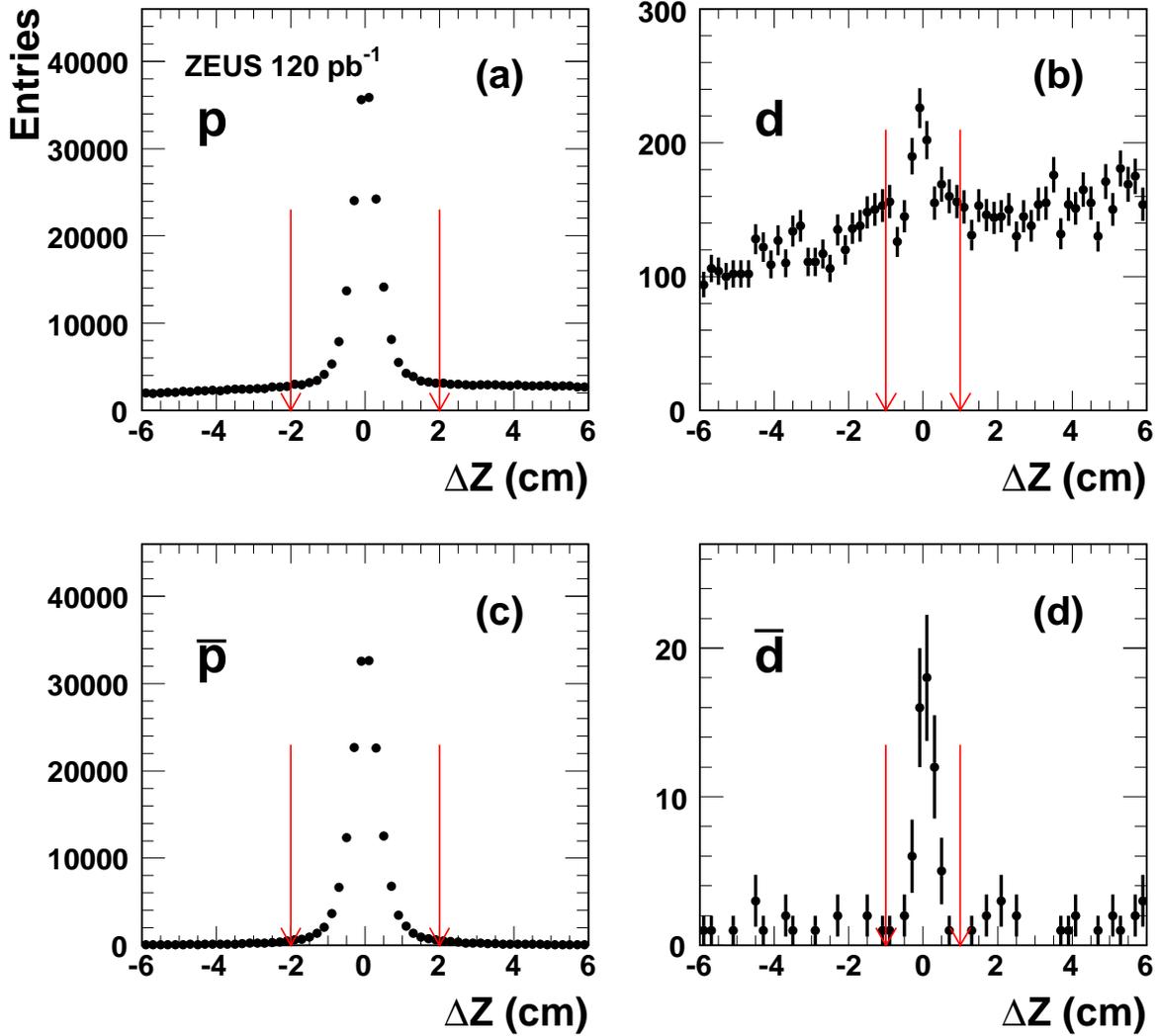}
\caption{The distributions of $\Delta Z$, the distance of  
the $Z$-component of the
track helix to $Z_{\mathrm{vtx}}$ 
for: (a)-(b)  particles and  (c)-(d) antiparticles, as indicated in the figure. 
The $p$, $\bar{p}$, $d$ and  $\bar{d}$  candidates 
were identified using the $dE/dx$ mass cuts (see text).
The arrows indicate the applied cuts. 
}
\label{dz}
\end{center}
\end{figure}

In order to
select $p$($\bar{p}$) and  $d$($\bar{d}$) originating from $ep$ collisions,
both $DCA$ and $\Delta Z$ cuts were removed and a statistical background subtraction based
on the $DCA$ distribution was performed.
The $\Delta Z$ distributions for $p$($\bar{p}$) and  $d$($\bar{d}$) after the mass cuts 
are shown in Figure~\ref{dz}. Clear peaks at $\Delta Z=0$ are observed.
To optimize the signal-over-background ratio for the $DCA$ distribution,
all candidates were selected  using the $\mid\Delta Z\mid <2(1)\cm$ restriction for $p,\bar{p}$ ($d,\bar{d}$).

\begin{figure}
\begin{center}
  \includegraphics[height=16.0cm]{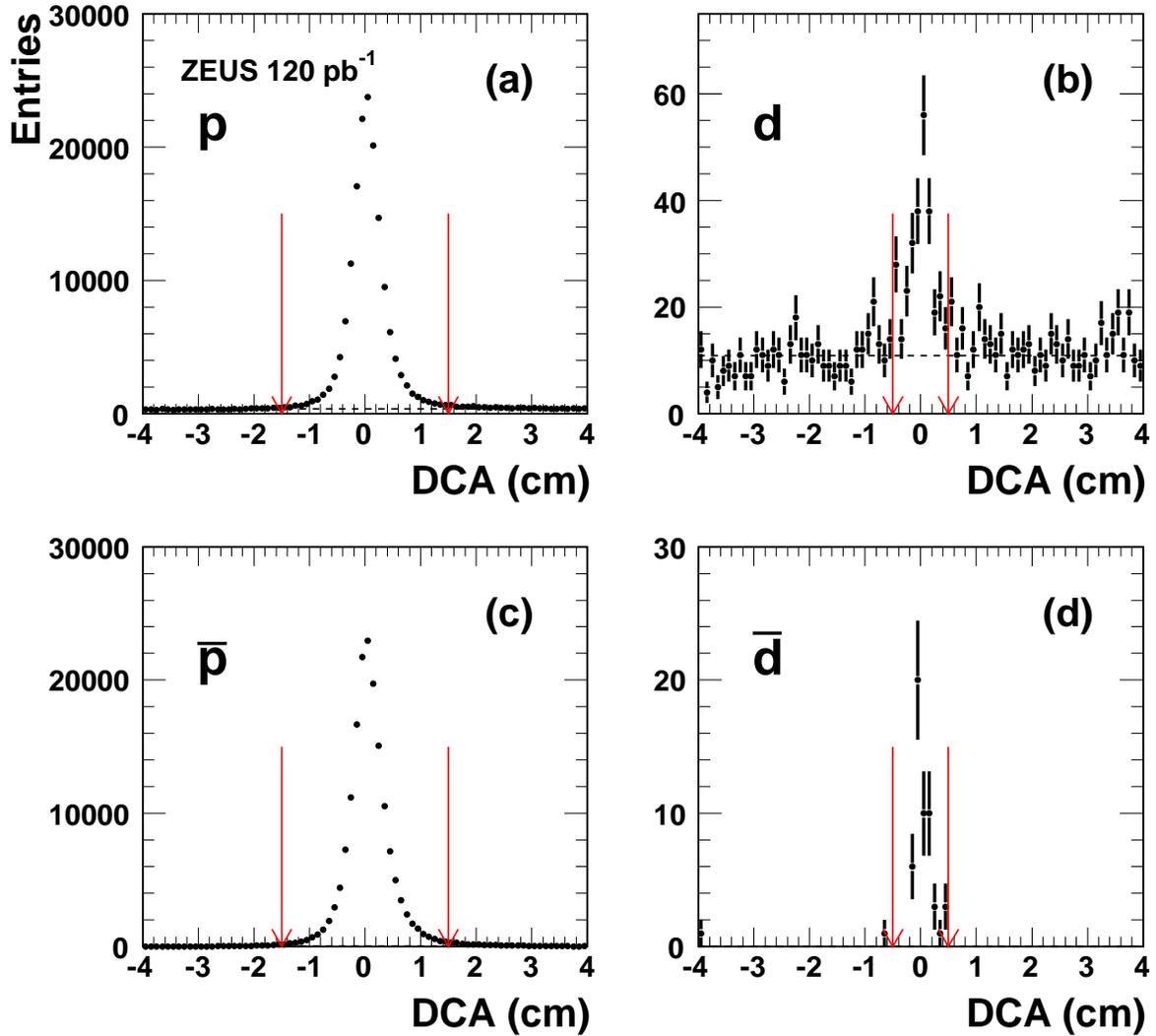}
\caption{The distributions of the distance of closest approach, DCA,  
for: (a)-(b)  particles and  (c)-(d) antiparticles.
The DCA are shown after the cut $\mid \Delta Z\mid <2(1)\cm$ as discussed in the text. 
The arrows indicate the signal region for the side-band background subtraction.
The dashed lines  show the fitted background level.
}
\label{dca}
\end{center}
\end{figure}

Figure~\ref{dca} shows the $DCA$ distributions
for $p$($\bar{p}$) and  $d$($\bar{d}$) candidates.
The distributions show peaks at zero due to tracks originating from the primary vertex. 
The number of particles originating from primary $ep$ collisions 
was determined using the side-band
background subtraction.
A linear fit to the $DCA$ distribution on either side of the peak region
in the range  $2<\mid DCA \mid <4\cm$  was performed. Then, the expected number
of background events in the signal region of $\mid DCA \mid <  1.5(0.5)\cm$ for $p,\bar{p}$ ($d,\bar{d}$) candidates
was subtracted.

The number of  $p$($\bar{p}$) obtained after the $DCA$ side-band
background subtraction
was  $1.52\times 10^5$ ($1.62\times 10^5$).
The numbers of $d$  and  $\bar{d}$ particles were
$177\pm 17$ and $53\pm 7$, respectively.
The difference in the observed numbers of $p$ and $\bar{p}$ can be 
explained by different $dE/dx$ efficiencies
and the mass cuts for positive and negative tracks.
Such a difference in the efficiencies for particles and antiparticles 
cannot explain the
difference in the observed numbers of $d$ and $\bar{d}$.

\begin{figure}
\begin{center}
  \includegraphics[height=16.0cm]{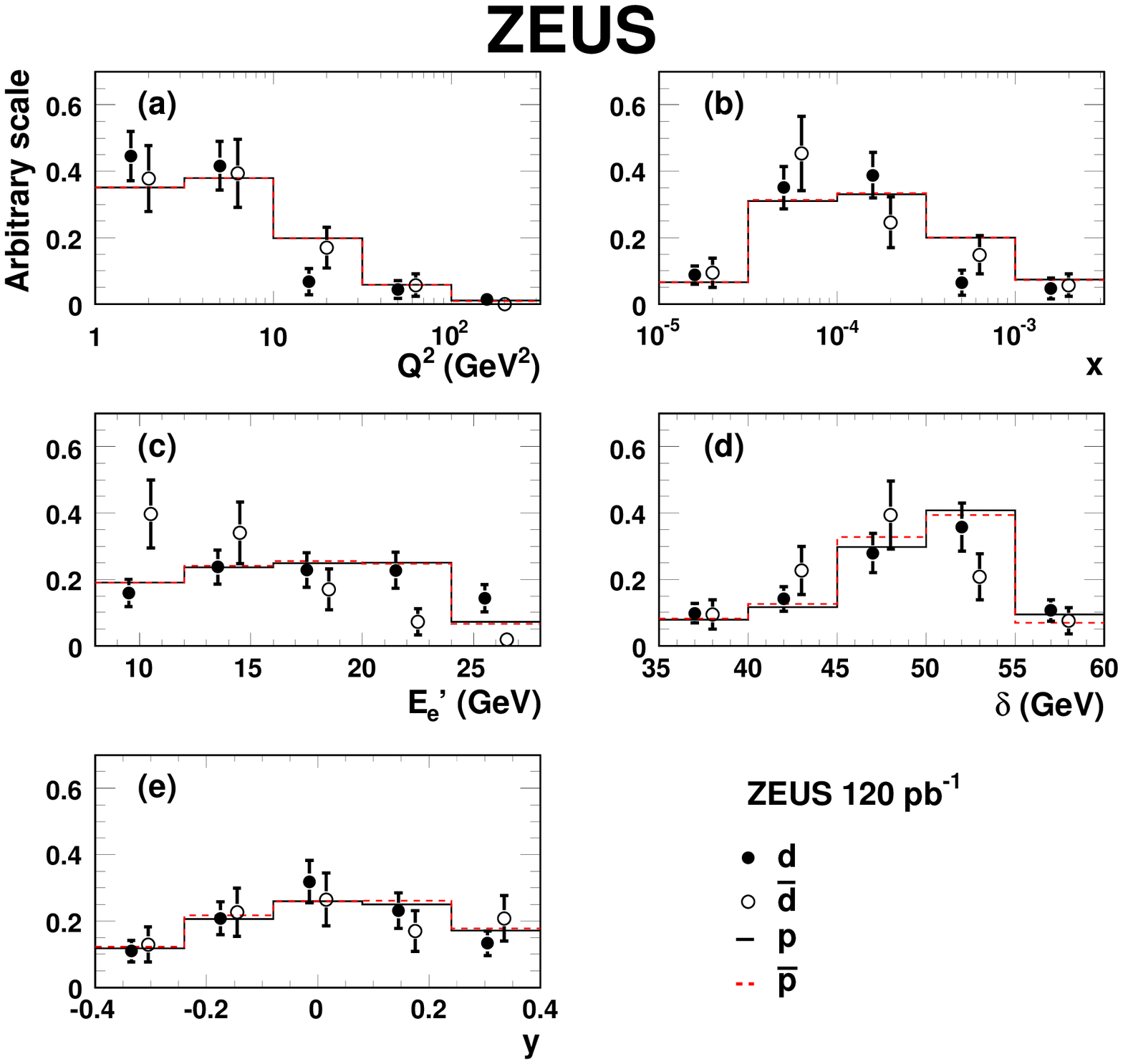}
\caption{The distributions of the number of events with at 
least one $d$($\bar{d}$) and $p$($\bar{p}$) 
candidate normalised to unity as a function of: (a)-(d)  DIS
kinematic variables and  (e)    
rapidity $y$. 
The points for $d$ and $\bar{d}$ 
are slightly shifted horizontally for clarity.
}
\label{dis_kin}
\end{center}
\end{figure}

Figure~\ref{dis_kin} shows the distributions  for
several DIS kinematic variables:  $Q^2_{e}$, $x_e$, $E_{e^{'}}$  and $\delta$.
In addition, rapidity ($y$) distributions for the selected candidates are shown.
The numbers of $p$($\bar{p}$) and  $d$($\bar{d}$)  candidates were  calculated
in each bin from the $DCA$  distributions
after the side-band background subtraction.
The distributions for $d$
are consistent with those for $p$ and $\bar{p}$, while 
the $\bar{d}$  sample shows some deviations
for the $E_{e^{'}}$ variable and, consequently, for the  $\delta$ variable.

\section{Studies of background processes}
\label{sec:bgas}

The following two background sources for heavy stable charged particles
were considered:
\begin{itemize}
\item[$\bullet$]
interactions of the proton (or electron) beam with residual gas in the beam pipe, 
termed beam-gas interactions;
\item[$\bullet$]
secondary interactions of particles in inactive material between the interaction point and
the central tracking detector.
\end{itemize}

\subsection{Beam-gas interactions}
\label{sec:bgas2}

The contribution from proton-gas interactions is  significantly reduced after
the ZEUS three-level trigger which requires
a scattered electron in the CAL.
In addition, the requirement to accept only events with more than three tracks
fitted to the primary vertex significantly diminishes the contribution from both
electron-gas and proton-gas events.
The remaining fraction of beam-gas
interactions can be assessed by studying the $Z_{\mathrm{vtx}}$ distribution.

\begin{figure}
\begin{center}
  \includegraphics[height=16.0cm]{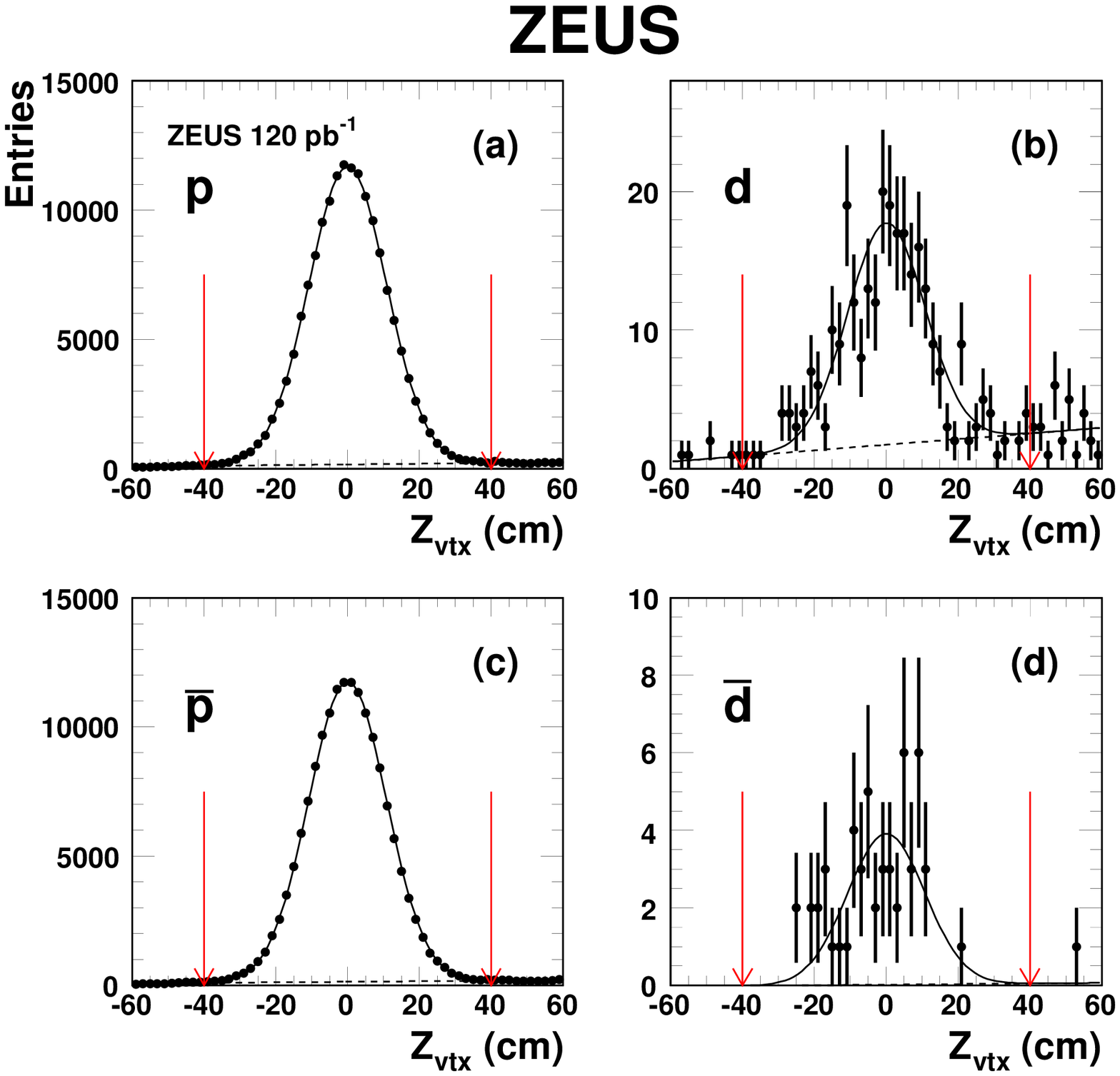}
\caption{The $Z_{\mathrm{vtx}}$ distributions 
for: (a)-(b)  particles and  (c)-(d) antiparticles, as indicated in the figure. 
The solid  lines show the  fit using a Gaussian distribution with a 
first-order polynomial function for the background description.
The dashed line shows the fitted background. 
The arrows indicate  the cuts applied for the final selection.
}
\label{dis_z}
\end{center}
\end{figure}

Figure~\ref{dis_z} shows the $Z_{\mathrm{vtx}}$ distributions
for events with at least one $p$($\bar{p}$)  or $d$($\bar{d}$)
candidate.
The distributions were  reconstructed
in the signal region $\mid\Delta Z\mid< 2(1)\cm$ and  $\mid DCA\mid<1.5(0.5)\cm$
for $p,\bar{p}$ ($d,\bar{d}$) candidates without the background subtraction.
Figure~\ref{dis_z} shows that there is essentially no beam-gas background for $\bar{d}$
events.
A small background for $d$ at positive $Z_{\mathrm{vtx}}$ is
expected from the DIS MC  generated
for inclusive DIS events in which $d$'s  are solely produced by secondary interactions in the material
in front of the CTD.
This background is expected to have a flat $DCA$  and, therefore, is subtracted by the
procedure described in Section~\ref{sec:identif}.  

The  $Z_{\mathrm{vtx}}$ distributions
were fitted using a Gaussian function with
a first-order polynomial for the background description.
The extracted Gaussian widths are fully consistent
with those  obtained for inclusive DIS events without the $d$ preselection.

To further study the $Z_{\mathrm{vtx}}$ distribution,
a special event selection was performed for non-colliding electron and proton bunches.
Since the requirement to detect an  electron with
energy $E_{e^{'}}\geq 8.5\gev$ significantly reduces the rate of such background events,
this requirement was not applied. All other tracking cuts were the same as in the $d$($\bar{d}$)  selection.
The requirement to accept events with at least three  tracks fitted to the primary vertex
rejects most of the beam-gas events ($\sim 95\%$ from the total number of the triggered events).
As expected, the remaining events show  clear peaks at zero for the $\Delta  Z$ and $DCA$ distributions, but
the reconstructed $Z_{\mathrm{vtx}}$ distribution did not show a peak at zero.

The enhancement at large  $Z_{\mathrm{vtx}}$ for $d$, 
which was found to be consistent with that originating from secondary interactions, could partially be
due to electron-gas interactions.
If one assumes that the background seen in Fig.~\ref{dis_z}(b)  is
due to non-$ep$ interactions, then the contribution from  
beam-gas interactions does not exceed $17\%$
of the total number of events with a deuteron.

\subsection{Secondary interactions on inactive material}
\label{sec:sinte}

A pure sample of DIS events will still contain deuterons
produced by secondary interactions of particles in material.
The aim of the side-band
background subtraction discussed in Sect.~\ref{sec:identif} was to remove such 
a background contribution,
assuming that the background processes do not create a residual peak at $\Delta Z=0$ and  $DCA=0$.
Several checks of this assumption are discussed below.

The $DCA$ and $\Delta Z$  distributions were investigated using a
MC simulation of inclusive DIS events without
$d(\bar{d})$  production at the generator level.
Deuterons from secondary interactions were selected as for the data.
The reconstructed $DCA$ and $\Delta Z$ for $d$ did not show a peak at zero.
A more detailed study of the $DCA$ and $\Delta Z$ distributions was possible
for $p$ not originating from  an $ep$  collision at the MC generator level,
since in this case the available MC statistics is significantly higher than for the $d$ case.
After the track-quality cuts,
no peak at zero was observed in the $DCA$ and $\Delta Z$ distributions.

If a deuteron is  produced by secondary interactions of the
particles from the DIS event in the surrounding matter,
the secondary $d$ will
not point precisely back to the interaction point,
and both $DCA$ and $\Delta Z$
distributions  will be wider than in case of $\bar{d}$
and $\bar{p}$.
Therefore, the $DCA$ and $\Delta Z$ distributions
were fitted with double-Gaussian distributions to establish the width
of the distributions. It was found that
the observed deuteron $DCA$ and $\Delta Z$ widths were consistent with
the corresponding widths for $p$ and $\bar{p}$.

One possible source for $d$ is the reaction
$N+N \to  d+\pi$, where one of the nucleons $N$ originates from $ep$ collision,
while the other one originates  from the detector material in front of the CTD.
For low initial nucleon momenta, the $DCA$ of the $d$ track is in general large
and it does not form an important background; at high initial nucleon momenta however,
the $DCA$ can become small enough that misidentification could become
important\footnote{Note that the  cross section for the reaction $N+N \to  d+\pi$
decreases rapidly with increasing energy.}.
Since the processes  $N+N \to  d+\pi$ can lead to an additional
charged pion, this source of background deuterons can be studied
by comparing the average charged  multiplicity
of tracks for  $d$  and $\bar{d}$ events.
In addition, the distance of closest approach,  $DCA12$,  between the $d$
track and other non-primary tracks in the same event
should have an enhancement at zero.
The study
indicated that the average number of tracks for $d$ events
is smaller than that for $\bar{d}$ events.
The rejection of events with  $|DCA12| <2\cm$
did not lead to a statistically significant reduction in 
the number of the observed $d$ events.

Secondary deuterons may also be produced in pickup ($p+n\to d$) reactions by
primary $p(n)$ interacting in the surrounding material. These deuterons,
peaking in the direction of the primary $p(n)$, point approximately
to  the interaction point and are therefore a potentially dangerous source
of background. Experimental data on the  pickup reactions at the relevant
energy are scarce and therefore only a rough estimate of the size of this
background is possible.
From the extrapolation of data on Sm$^{154}$~\cite{blasi} 
and $C$~\cite{Ero:1987gm,PhysRevC.30.593} targets using
the K.~Kikuchi theory~\cite{kuki}  to allow for the change
of material,
the estimated $d$ background from the pickup
reaction was in the range $1-10\%$ of the total number of observed $d$ events,
depending on the extrapolation input.

The angular distributions of $d$ from pickup
reactions have also been investigated in several experiments \cite{roos,Fager,Ero:1987gm}
for various targets and for a range of $p/M$ similar to the present analysis.
In all cases, the angular distribution
of $d$ observed in these experiments would lead to a much wider $DCA$ than that
shown in  Figure~\ref{dca}(b).


\section{Detector corrections}
\label{sec:corrections}

In this analysis,
all measurements are based on event ratios, therefore, the 
detector corrections due to DIS event selection and trigger efficiency were found to be small
and thus are not discussed here.
The detector corrections for the tracking efficiency
and the efficiency of the $dE/dx$ cuts are described below.

\subsection{Tracking efficiency}
\label{st:eff}

The efficiency due to the track reconstruction, $\varepsilon$,  was
estimated separately for $p$ ($\bar{p}$) and $d$  using the {\sc Ariadne} MC
model (with $d$ included at the generator level).
The obtained efficiencies
are about  0.95 for $p$ and $d$ and
0.90 for $\bar{p}$.

The method cannot be applied to $\bar{d}$
which are not treated in the  {\sc Geant}  simulation.
Therefore, the tracking efficiency for $\bar{d}$ was modelled as
$\varepsilon (\bar{d}) = \varepsilon (d)  \varepsilon (\bar{p}) / \varepsilon (p)$.
In the expression above,
the hit reconstruction efficiency is described by the first term, $\varepsilon (d)$,
while the absorption loss (including annihilation) of $\bar{d}$ and $\bar{p}$ are assumed
to be similar. This modelling assumes that the cross sections of annihilation
in the detector material are the same for $\bar{d}$ and $\bar{p}$,
since the inelastic nuclear cross section of $\bar{p}$ is much larger than that of
$\bar{n}$ for the momentum region less than  $0.4$ GeV~\cite{xsec}.
The use of the geometrical model discussed in \cite{phenix1,xsec} and the model in which 
the $\bar{p}$ and $\bar{n}$ inelastic absorption cross sections are added linearly~\cite{aleph_deu,xsec} to obtain 
the inelastic nuclear cross section of $\bar{d}$,
reduces  $\varepsilon (\bar{d})$  by $1\%$ and $5\%$, respectively.

\subsection{Efficiency of the  $\mathbf{dE/dx}$  cuts}

Another important contribution to the efficiency comes from the $dE/dx$ threshold cuts  and
the mass cuts.
The inefficiency due to the $dE/dx$ requirements
were estimated separately for positive and negative tracks
using $\Lambda \to p \pi$ (+c.c.) decays. In this approach,
protons were identified from the $\Lambda$ peak and then the proton
$dE/dx$ selection efficiency was reconstructed as the ratio of the events without and with
the $dE/dx$ requirement.
These efficiencies were determined as a function of $p/M$.
The efficiency for each $p_T/M$ bin was corrected by reweighting
the $p/M$ distributions using {\sc Ariadne}.
The average  efficiency of the $dE/dx$ cuts for $d(\bar{d})$ is $0.7$ for $p_T/M < 0.5$.
For larger momenta, the  efficiency decreases due to the 
$dE/dx>2.5$~mips cut. The signal extraction is not possible for $p_T/M > 0.7$ due to
a very small efficiency.
For the low-momentum region $p_T/M<0.5$,
the efficiencies for negative tracks tend to be
larger than for positive tracks. The $dE/dx$ efficiency for $p(\bar{p})$ is higher by $15\%$
than that for $d(\bar{d})$.

Alternatively, the overall tracking and the  $dE/dx$ efficiency was calculated
using the {\sc Ariadne} MC model; consistent results with the approach discussed above
were  found.

\section{Systematic uncertainties}
\label{sec:sys}

The systematic uncertainties
were evaluated by
changing the selection and the analysis procedure.
Only the largest contribution of each cut variation  for the  final invariant cross section is
given below. The following sources of systematic uncertainties were studied:

\begin{itemize}

\item[$\bullet$]
efficiency of the  track reconstruction and selection.
The systematic  uncertainty on  the tracking efficiency for $p$, $\bar{p}$,
$d$ was  $\pm 2\%$.
This systematic uncertainty was found after variations of the track-quality cuts.
For $\bar{d}$, the systematic
uncertainty, $\pm 5\%$, includes both
the effect of track-quality-cut variations  and the reduction in  $\varepsilon (\bar{d})$
when the linear model for the $\bar{d}$ absorption was  used
(see Section~\ref{st:eff}); 

\item[$\bullet$]
efficiency due to the $dE/dx$ selection. This systematic uncertainty 
was estimated by varying the cut $dE/dx>2.5$ mips within the $dE/dx$
resolution and by using the MC simulation. 
This systematic  uncertainty was $\pm 5\%$. 
For the lowest $p_T/M$ bin, the uncertainty was $\pm 10\%$;

\item[$\bullet$]
variations in the particle yields associated with the signal extraction:

\begin{itemize}
\item  the  number of $d$($\bar{d}$) were reconstructed using
a Gaussian fit to the $DCA$ distribution with a first-order polynomial for the
background description;

\item
the region used to determine the background
for the  side-band background subtraction was reduced to $1.5<\mid DCA\mid <3.5\cm$;

\item the $DCA$ cut for the side-band background subtraction was varied within
its resolution of $\pm 0.1\cm$;

\item for the side-band background subtraction,
the background shape
was taken from the MC (without $d$ at the generator level);

\item the cut on $\Delta Z$ was varied by $\pm 0.2\cm$;
\end{itemize}

These variations lowered the production yields by 
$5.0\%$ for $p$,
$2.2\%$ for $\bar{p}$,
$26.0\%$ for $d$ and $6.1\%$ for $\bar{d}$.
The largest effect originates from the conservative treatment
of the shape of the $DCA$ background.
The upper systematic error was below $1\%$ for $p$, $\bar{p}$ and $\bar{d}$,
and $11\%$ for $d$.

\item[$\bullet$]
the background contribution under the $Z_{\mathrm{vtx}}$ peak for
$d$  events
was assumed to be due to beam-gas interactions and,
therefore, it was subtracted ($-4\%$ contribution for
$p$, $\bar{p}$,  $\bar{d}$ and $-17\%$ contribution for $d$);

\item[$\bullet$]
the correction for $\Lambda$ decays applied for the $p$($\bar{p}$) sample
was changed by $\pm 10\%$
(see Section~\ref{sec:xcross}). The size of this uncertainty, which is similar to that in other
publications~\cite{h1deuterons,aleph_deu},  was determined by the
uncertainty on the strangeness suppression factor
in  the {\sc Ariadne} model;

\item[$\bullet$]
variations of the DIS-selection cuts.
The cut on the energy of the scattered electron was increased to $10\gev$,
and the lower cut on the $\delta$ distribution was tightened to $40\gev$.
The cut on $Z_{\mathrm{vtx}}$ was varied by $\pm 5\cm$.
The cut on the number of primary tracks was increased from three to four.
These variations led to changes of 
$^{+3.3}_{-4.1}\%$ for $p$, $^{+3.6}_{-4.4}\%$ for $\bar{p}$,
$^{+3.7}_{-8.5}\%$ for $d$ and $^{+5.7}_{-13.3}\%$ for $\bar{d}$.  
Variations of the cuts on
$y_{e}$ and $y_{\mathrm{JB}}$ distributions showed a negligible  effect.

\end{itemize}

The overall systematic uncertainty was  determined by adding
the  above  uncertainties in quadrature. The largest experimental
uncertainty was due to the uncertainties on the tracking efficiency and the signal extraction.

\section{Results}
\label{sec:results}

\subsection{Production cross sections and $B_2$}
\label{sec:xcross}

For each particle type $i$, the invariant differential cross section 
can be
calculated from the rapidity range $\Delta y$ and the
transverse momentum $p_{T,i}$  of a corresponding particle through  
$$
\frac{\gamma_i}{\sigma_{tot}} \frac{\ud^3 \sigma_i} {\ud (p_i/M_i)^3} =
\frac{1}{ N_{\mathrm{DIS}}} \frac{1}{2\pi (p_{T,i}/M_i) \Delta y} \frac{N_i} {\Delta (p_{T,i}/M_i)},
$$
where the subscript
$i$ denotes a $p$($\bar{p})$ or a $d(\bar{d})$, 
$N_i$ is the particle yield in each $p_{T,i}/M_i$ bin after the correction
for the tracking efficiencies and the particle selection and
$N_{\mathrm{DIS}} = 2.59 \times 10^7$ is the number of DIS events used in the analysis.
For the present measurement, $\Delta y=0.8$
and $\Delta(p_{T,i}/M_{i}) = 0.1$ are the bin sizes.
For comparisons with other experiments, the $p$($\bar{p}$) rate
was corrected for the decay products of $\Lambda$.
A  correction factor of  $0.79$  was  estimated from the {\sc Ariadne} simulation
which gives an adequate description of $K_S^0$ and $\Lambda$ production \cite{2006wz}.

\begin{figure}
\begin{center}
 \includegraphics[height=16.0cm]{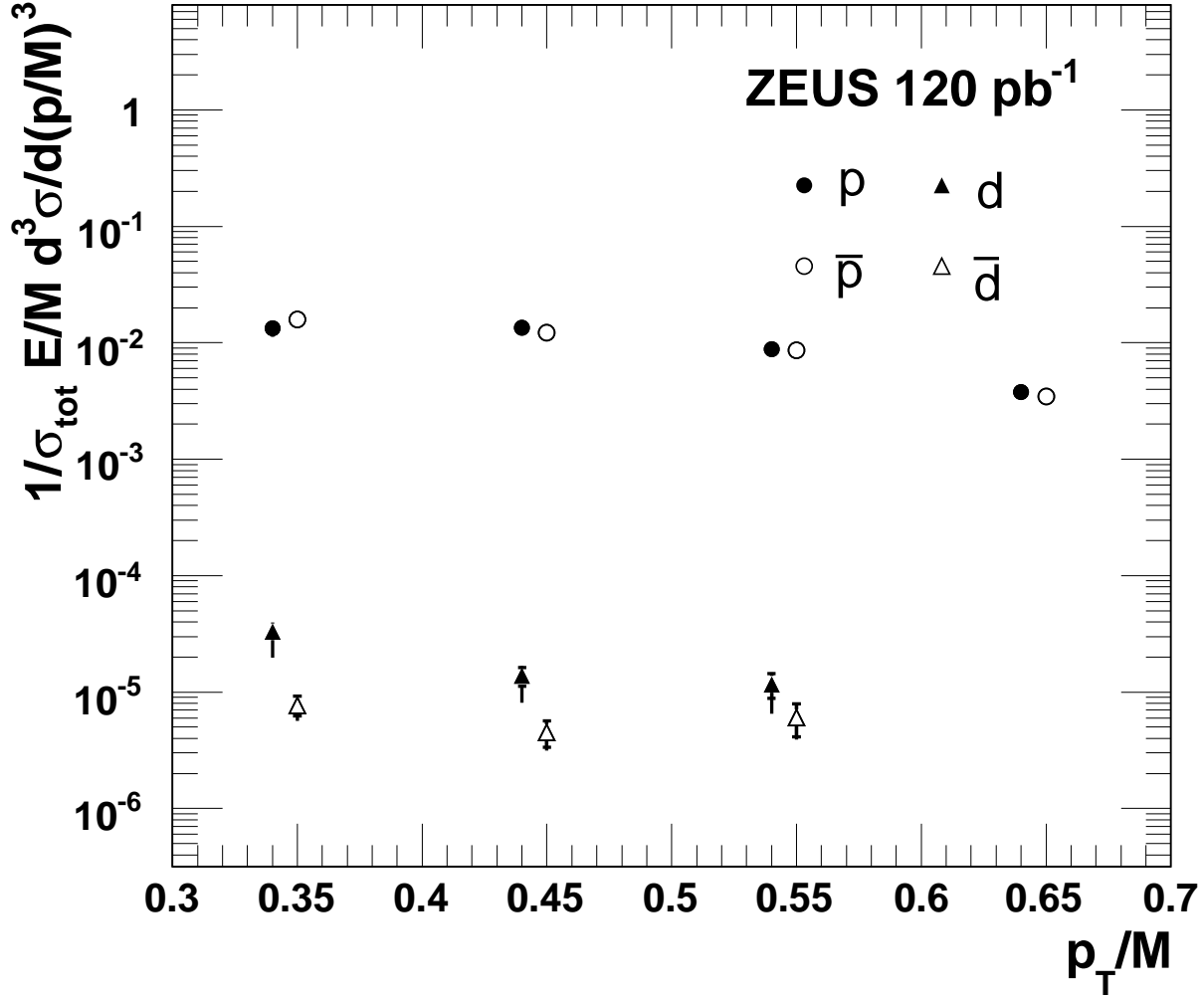}
\caption{The invariant differential cross sections 
for $p(\bar{p})$ and  $d(\bar{d})$  produced in DIS $ep$ collisions
as  a function of $p_T/M$.
The inner error bars show the statistical uncertainties,
the outer ones show statistical and systematic uncertainties added
in quadrature.
For clarity, the points for particles and antiparticles
are slightly shifted horizontally with respect to the corresponding $p_T/M$.
}
\label{xcross}
\end{center}
\end{figure}

\begin{figure}
\begin{center}
  \includegraphics[height=16.0cm]{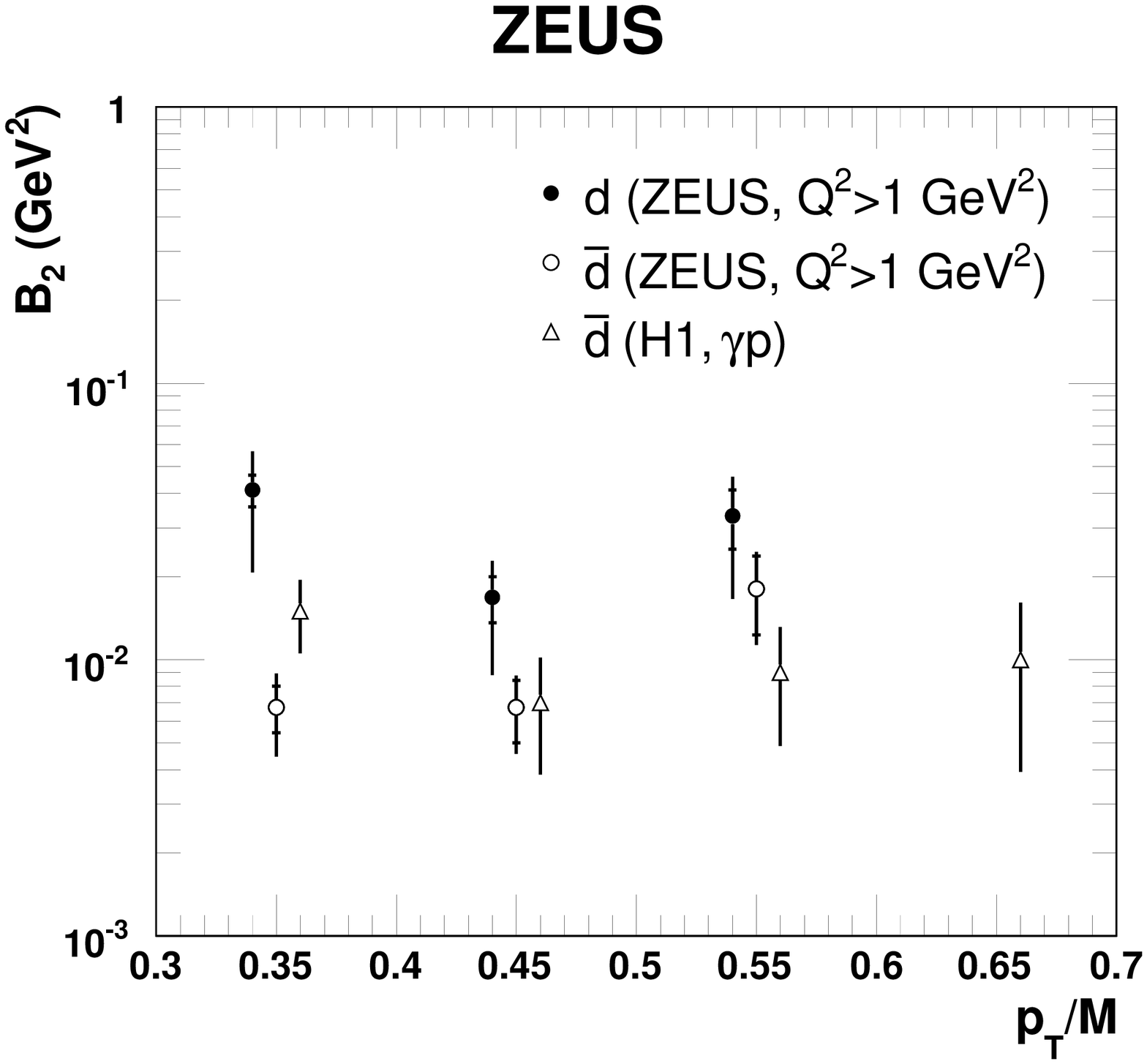}
\caption{The $p_T/M$ dependence of the parameter $B_2$ for
$d$  and $\bar{d}$ produced in DIS $ep$ collisions and in photoproduction~\protect\cite{h1deuterons}.
The inner error bars show the statistical uncertainties,
the outer ones show statistical and systematic uncertainties added
in quadrature.
For clarity, the points for particles and antiparticles
are slightly shifted horizontally with respect to the corresponding $p_T/M$.
}
\label{b2}
\end{center}
\end{figure}

The invariant
differential cross sections as a function of $p_T/M$ for
$p$($\bar{p}$) and $d$($\bar{d}$) are shown in Fig.~\ref{xcross} and 
given in Tables~\ref{tab1} and~\ref{tab2}.
The $d$($\bar{d}$) invariant cross section is smaller by approximately three  orders of magnitude
than that of $p(\bar{p})$.
These cross sections were used to extract the coalescence parameter $B_2$ as discussed in Section~\ref{sec-dec}.
The parameter $B_2$ is shown in Fig.~\ref{b2} and
listed in Tables~\ref{tab3} and~\ref{tab4}.
For $d$,  $B_2$ tends to be  higher than for $\bar{d}$, especially at low $p_T/M$.
The value of $B_2$ for $\bar{d}$
is in agreement with the measurements in photoproduction~\cite{h1deuterons}, but
larger than that observed in $e^+e^-$ annihilation at the $Z$ resonance~\cite{aleph_deu}.
The measured $B_2$ is also significantly larger than that observed in heavy-ion
collisions~\cite{phenix1}.

The events containing at least one $p$($\bar{p}$) or
$d$($\bar{d}$) were analysed in the Breit
frame \cite{feynman:1972:photon, *zpf:c2:237}.
The number of events with
$p$($\bar{p}$) in the current region
of the Breit frame
was about $2.5\%$ of  the total number of observed events with $p$($\bar{p}$).
In this region, neither $d$ nor $\bar{d}$ was found.
Since the current region of the Breit frame is analogous to a single
hemisphere of $e^+e^-$, the observation of $d$($\bar{d}$) reported in this paper
is not in contradiction with the low $\bar{d}$ rate observed in
$e^+e^-$~\cite{argus1,*argus2,opal1,aleph_deu}.

\subsection{Production ratios}
\label{sec:ratio}

\begin{figure}
\begin{center}
  \includegraphics[height=16.0cm]{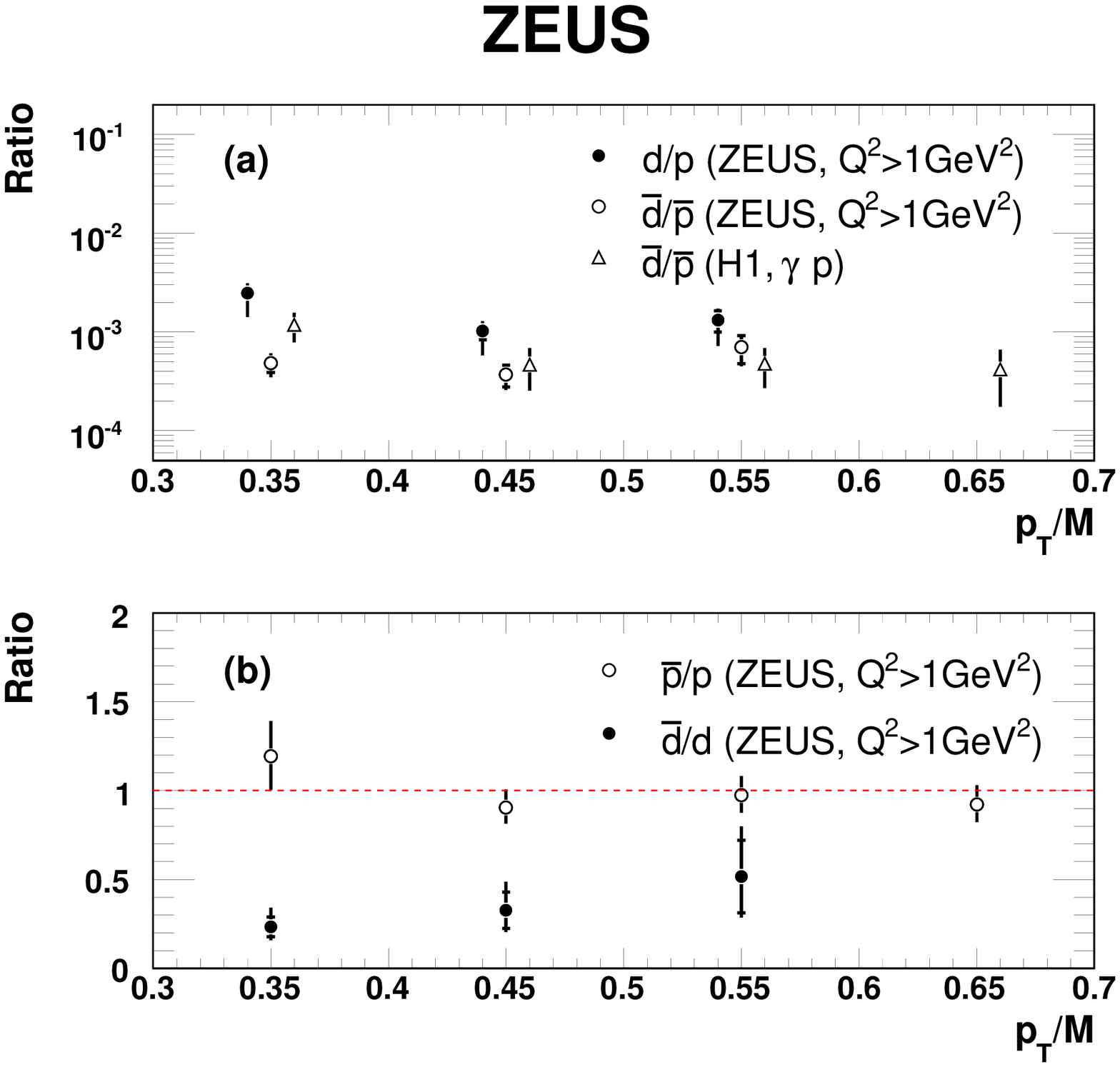}
\caption{(a) $d/p$ and
$\bar{d}/ \bar{p}$ production ratios as a function of $p_T/M$ 
compared to the H1 photoproduction results~\protect\cite{h1deuterons}.    
(b) the $\bar{d}/ d$ and $\bar{p}/ p$ production ratios as a function of $p_T/M$.
The inner error bars show the statistical uncertainties,
the outer ones show statistical and systematic uncertainties added
in quadrature. 
The points in (a) 
are slightly shifted horizontally for clarity.
}
\label{dd2}
\end{center}
\end{figure}

The detector-corrected  $d/p$ and $\bar{d}/\bar{p}$
ratios as a function of $p_T/M$ are shown in Fig.~\ref{dd2}(a) and listed in 
Tables~\ref{tab3} and \ref{tab4}.
For the antiparticle ratio,
there is a good agreement with the H1 published data for photoproduction
\cite{h1deuterons}, as well as with $pp$ data \cite{pp1,*pp2}.
A similar $\bar{d}/\bar{p}$ ratio was also observed in
hadronic $\Upsilon (1S)$ and $\Upsilon (2S)$ decays
\cite{argus1,*argus2}.

The $\bar{d}/d$ and $\bar{p}/p$ ratios as a function of
$p_T/M$ are shown in Fig.~\ref{dd2}(b) and listed in Table~\ref{tab5}.
The $\bar{p}/p$ ratio is consistent with unity, as 
expected from
hadronisation of quark  and gluon jets.
The dominant uncertainty on the ratio is due to systematic effects
associated with the track selection and reconstruction.

The production rate of $d$ is higher than that of  $\bar{d}$, especially
at low $p_T$.
Under the assumption that secondary interactions do not produce an
enhancement at $DCA=0$ for the $d$ case, the result would indicate
that the relation between
$\bar{d}/d$ and $(\bar{p}/p)^2$ expected from the coalescence model does not hold
in the central fragmentation region of $ep$ DIS collisions.

For collisions involving incoming baryon beams,  there are
several
models~\cite{Garvey:2000dk,*Chekanov:2005wf,*Bopp:2004qz,*Bopp:2006bp,Kopeliovich:1996qb,*Kopeliovich:1998ps}
that predict baryon-antibaryon production asymmetry in the central rapidity region.
A $p-\bar{p}$ asymmetry in proton-induced reactions
is predicted to be as high as $7\%$~\cite{Kopeliovich:1996qb,*Kopeliovich:1998ps}.
Given the experimental uncertainty, this measurement is not sensitive
to the expected small $p-\bar{p}$ asymmetry.

In heavy-ion collisions,
the $\bar{d}$ to $d$ production ratio is expected to be smaller than unity~\cite{Leupold}.
A recent measurement at RHIC~\cite{phenix1}
indicated  a lower production rate of  $\bar{d}$  compared to that of $d$. 
The average value of the ratio $\bar{d}/d=0.47\pm 0.03$ 
was compatible with the square of the $\bar{p}/p=0.73\pm 0.01$ ratio. 
Assuming the same size of the production volume for baryons and antibaryons, 
this RHIC result is consistent with the coalescence model.
A similar conclusion
was obtained earlier in fixed-target $pp$~\cite{pp1,*pp2} and $pA$~\cite{pA3} experiments.
For $e^+e^-$ collisions,
the $d$ yield is compatible with that of $\bar{d}$  within
the large uncertainties~\cite{aleph_deu,cleo_deu}.

\section{Summary}
\label{sec:summary}

The first observation of $d(\bar{d})$ in $ep$ collisions in the DIS regime
at HERA is presented.
The production rate of $d(\bar{d})$ is smaller than that 
for  $p(\bar{p})$ by three orders of magnitude,
which is in broad agreement with other experiments.

The production of $d(\bar{d})$  was studied in terms of the coalescence model.
The coalescence parameter
is in agreement with the measurements in photoproduction at HERA. However, it
is larger than that measured in $e^+e^-$
annihilation at the $Z$ resonance.

The production rate of $p$ is consistent with that of $\bar{p}$  in the
kinematic range $0.3<p_T/M<0.7$. Due to significant uncertainties,
it is not possible to test models that predict a small baryon-antibaryon asymmetry
in the central fragmentation region.

For the same kinematic region,
the production rate of $d$ is
higher than that for $\bar{d}$.
If the observed $d$
are solely attributed to deuterons produced in primary $ep$ collisions, the results
would indicate that the coalescence model with the same source volume for $d$ and  
$\bar{d}$ cannot fully explain the production of $d(\bar{d})$ in DIS.

\section*{Acknowledgements}
\vspace{0.3cm}
We thank the DESY Directorate for their strong support and encouragement.
The remarkable achievements of the HERA machine group were essential for
the successful completion of this work and are greatly appreciated. We
are grateful for the support of the DESY computing and network services.
The design, construction and installation of the ZEUS detector have been
made possible owing to the ingenuity and effort of many people from DESY
and home institutes who are not listed as authors.
We thank Prof. D.~Heinz and Prof. T.~Sloan for the useful discussion of this topic.

\vfill\eject

{
\def\bibname{\Large\bf References}
\def\refname{\Large\bf References}
\pagestyle{plain}
\ifzeusbst
  \bibliographystyle{./BiBTeX/bst/l4z_default}
\fi
\ifzdrftbst
  \bibliographystyle{./BiBTeX/bst/l4z_draft}
\fi
\ifzbstepj
  \bibliographystyle{./BiBTeX/bst/l4z_epj}
\fi
\ifzbstnp
  \bibliographystyle{./BiBTeX/bst/l4z_np}
\fi
\ifzbstpl
  \bibliographystyle{./BiBTeX/bst/l4z_pl}
\fi
{\raggedright
\bibliography{./BiBTeX/user/syn.bib,%
              ./BiBTeX/bib/l4z_articles.bib,%
              ./BiBTeX/bib/l4z_books.bib,%
              ./BiBTeX/bib/l4z_conferences.bib,%
              ./BiBTeX/bib/l4z_h1.bib,%
              ./BiBTeX/bib/l4z_misc.bib,%
              ./BiBTeX/bib/l4z_old.bib,%
              ./BiBTeX/bib/l4z_preprints.bib,%
              ./BiBTeX/bib/l4z_replaced.bib,%
              ./BiBTeX/bib/l4z_temporary.bib,%
              ./BiBTeX/bib/l4z_zeus.bib}}
}
\vfill\eject

%

\begin{table}
\centering

\begin{tabular}{|c|c|c|c|}
\hline
$p_{T} / M$ &   $( \gamma_p /\sigma_{tot}) \ud^3\sigma_p /\ud (p_p/M_p)^3 (\times 10^{-2})$  & 
$( \gamma_d / \sigma_{tot}) \ud^3\sigma_d /\ud (p_d/M_d)^3 (\times 10^{-5})$  \\
\hline\hline 
$0.3~\text{--}~0.4$  & $1.33\pm0.01^{+0.19}_{-0.21}$   & $3.29\pm 0.43^{+0.50}_{-1.24}$   \\ \hline
$0.4~\text{--}~0.5$  & $1.34\pm0.01^{+0.16}_{-0.18}$  & $1.37\pm 0.26^{+0.17}_{-0.51}$ \\ \hline
$0.5~\text{--}~0.6$  & $0.88\pm0.01^{+0.10}_{-0.12}$    & $1.16\pm 0.28^{+0.14}_{-0.42}$  \\ \hline
$0.6~\text{--}~0.7$  & $0.38\pm0.01^{+0.04}_{-0.05}$   &   \text{-----}   \\ \hline
\end{tabular}

\caption{
The measured
invariant cross sections for the production of $p$ and  $d$ in DIS
as a function of $p_T/M$. The statistical and systematic uncertainties are also listed.  
}
\label{tab1}
\end{table}

\begin{table}
\centering
\begin{tabular}{|c|c|c|c|}
\hline
$p_{T} / M$ &   $( \gamma_{\bar p} /\sigma_{tot}) \ud^3\sigma_{\bar p} /\ud (p_{\bar p}/M_{\bar p})^3 (\times 10^{-2})$  &
$( \gamma_{\bar d} / \sigma_{tot}) \ud^3\sigma_{\bar d} /\ud (p_{\bar d}/M_{\bar d})^3 (\times 10^{-5})$  \\
\hline\hline
$0.3~\text{--}~0.4$  & $1.59\pm 0.01^{+0.16}_{-0.19}$   & $0.77\pm0.15^{+0.09}_{-0.14}$   \\ \hline
$0.4~\text{--}~0.5$  & $1.21\pm 0.01^{+0.07}_{-0.09}$  & $0.45 \pm 0.11^{+0.03}_{-0.07}$ \\ \hline
$0.5~\text{--}~0.6$  & $0.86\pm 0.01^{+0.05}_{-0.07}$    & $0.60\pm 0.19^{+0.05}_{-0.09}$  \\ \hline
$0.6~\text{--}~0.7$  & $0.35\pm 0.01^{+0.02}_{-0.03}$   &  \text{-----}   \\ \hline 
\end{tabular}

\caption{
The measured
invariant cross sections for the production of $\bar{p}$ and  $\bar{d}$ in DIS
as a function of $p_T/M$.
The statistical and systematic uncertainties are also listed.
}
\label{tab2}
\end{table}

\begin{table}
\centering
\begin{tabular}{|c|c|c|c|}
\hline
$p_{T} / M$  & $R(d/p) (\times 10^{-3})$  & $B_{2} (d) (10^{-2} \mathrm{GeV}^{2})$ \\
\hline\hline
$0.3~\text{--}~0.4$  & $2.48\pm0.33^{+0.55}_{-1.00}$  &  $4.11\pm 0.54^{+1.47}_{-1.97}$  \\ \hline
$0.4~\text{--}~0.5$  & $1.02\pm0.19^{+0.19}_{-0.40}$  &  $1.68\pm 0.32^{+0.50}_{-0.74}$  \\ \hline
$0.5~\text{--}~0.6$  & $1.32\pm0.32^{+0.24}_{-0.51}$  &  $3.31\pm 0.80^{+0.99}_{-1.45}$ \\ \hline
$0.6~\text{--}~0.7$  & \text{-----}   &   \text{-----}  \\ \hline \hline 
$0.3~\text{--}~0.7$  & $1.88\pm0.20^{+0.40}_{-0.75}$  &   $3.32 \pm0.34^{+1.13}_{-1.55}$ \\ 
\hline
\end{tabular}

\caption{
The measured $d$-to-$p$  production ratio
and the parameter $B_2$ for $d$  as a function
of $p_T/M$. The last row of the table shows the data in the full measured
phase space.   
The statistical and systematic uncertainties are also listed.
}
\label{tab3}
\end{table}

\begin{table}
\centering
\begin{tabular}{|c|c|c|c|}
\hline
$p_{T} / M$  & $R(\bar{d}/\bar{p}) (\times 10^{-3})$  & $B_{2} (\bar{d}) (10^{-2} \mathrm{GeV}^{2})$ \\
\hline\hline
$0.3~\text{--}~0.4$  & $0.48\pm0.09^{+0.08}_{-0.10}$  &  $0.67\pm 0.13^{+0.18}_{-0.19}$  \\ \hline
$0.4~\text{--}~0.5$  & $0.37\pm0.09^{+0.04}_{-0.06}$  &  $0.67\pm 0.17^{+0.12}_{-0.13}$  \\ \hline
$0.5~\text{--}~0.6$  & $0.70\pm0.22^{+0.08}_{-0.12}$  &  $1.80\pm 0.57^{+0.31}_{-0.36}$ \\ \hline
$0.6~\text{--}~0.7$  & \text{-----}          &  \text{-----}  \\ \hline \hline
$0.3~\text{--}~0.7$  & $0.49\pm0.07^{+0.07}_{-0.09}$  &  $0.89\pm0.14^{+0.19}_{-0.20}$ \\ 
\hline
\end{tabular}

\caption{
The measured $\bar{d}$-to-$\bar{p}$  production ratio
and the parameter $B_2$ for $\bar{d}$ as a function
of $p_T/M$. The last row of the table shows the data in the full measured
phase space. The statistical and systematic uncertainties are also listed. 
}
\label{tab4}
\end{table}

\begin{table}
\centering
\begin{tabular}{|c|c|c|}
\hline
$p_{T} / M$  & $R(\bar{p}/p)$ &  $R(\bar{d}/d)$  \\
\hline\hline
$0.3~\text{--}~0.4$  & $1.19\pm 0.01^{+0.20}_{-0.19}$  & $0.23 \pm 0.05^{+0.09}_{-0.05}$ \\ \hline
$0.4~\text{--}~0.5$  & $0.90\pm 0.01^{+0.10}_{-0.09}$  & $0.33\pm 0.10^{+0.12}_{-0.07}$ \\ \hline
$0.5~\text{--}~0.6$  & $0.97\pm 0.01^{+0.11}_{-0.10}$  & $0.52\pm 0.21^{+0.19}_{-0.10}$ \\ \hline
$0.6~\text{--}~0.7$  & $0.92\pm 0.03^{+0.10}_{-0.09}$  & \text{-----}  \\ \hline\hline 
$0.3~\text{--}~0.7$  & $1.05\pm 0.01^{+0.15}_{-0.14}$  & $0.31\pm 0.05^{+0.11}_{-0.06}$ \\ 
\hline
\end{tabular}
\caption{
The measured $\bar{p}$-to-$p$  and $\bar{d}$-to-$d$  production ratios
as a function
of $p_T/M$. The last row of the table shows the data in the full measured
phase space.
The statistical and systematic uncertainties are also listed.
}
\label{tab5}
\end{table}

%

%
\end{document}